
%
\let\includefigures=\iftrue
%
%
%
\input harvmac.tex
\includefigures
\message{If you do not have epsf.tex (to include figures),}
\message{change the option at the top of the tex file.}
\input epsf
\epsfclipon
\def\fig#1#2{\topinsert\epsffile{#1}\noindent{#2}\endinsert}
\else
\def\fig#1#2{}
\fi
\def\Title#1#2{\rightline{#1}
\ifx\answ\bigans\nopagenumbers\pageno0\vskip1in%
\baselineskip 15pt plus 1pt minus 1pt
\else
\def\listrefs{\footatend\vskip 1in\immediate\closeout\rfile\writestoppt
\baselineskip=14pt\centerline{{\bf References}}\bigskip{\frenchspacing%
\parindent=20pt\escapechar=` \input
refs.tmp\vfill\eject}\nonfrenchspacing}
\pageno1\vskip.8in\fi \centerline{\titlefont #2}\vskip .5in}

\ifx\answ\bigans\def\tcbreak#1{}\else\def\tcbreak#1{\cr&{#1}}\fi
%

\def\CL {{\cal L}}

\def\inbar{\,\vrule height1.5ex width.4pt depth0pt}
\def\IB{\relax{\rm I\kern-.18em B}}
\def\IC{\relax\hbox{$\inbar\kern-.3em{\rm C}$}}
\def\IP{\relax{\rm I\kern-.18em P}}
\def\IR{\relax{\rm I\kern-.18em R}}

\def\tr{{\rm Tr}}
\def\pam{\partial_{-}}
\def\pap{\partial_{+}}
\def\sgn{{\rm sgn~}}

\def\half{{1\over 2}}

\def\th{\theta}

\def\ep{\epsilon}

\def\D{\Delta}

\def\pa{\partial}

\def\la{\langle}
\def\ra{\rangle}
\def\pa{\partial}
\def\e{{\rm e}}

\def\Dslash{\rlap{\hskip0.2em/}D}
\def\su{$SU(N)$\ }
\def\mn{{\mu\nu}}
\def\emn{{\ep^{\mu\nu}}}
\Title{\vbox{\baselineskip12pt
\hfill{\vbox{
\hbox{LPTENS-94/2}\hbox{IASSNS-HEP-94/3\hfil}\hbox{RU-94-8\hfil}}}}}
{\vbox{\centerline{Generalized Two-Dimensional QCD}}}
\centerline{Michael R. Douglas}
\smallskip
\centerline{Dept. of Physics and Astronomy}
\centerline{Rutgers University }
\centerline{\tt mrd@physics.rutgers.edu}
\bigskip
\centerline{Keke Li}
\smallskip
\centerline{Institute for Advanced Study}
\centerline{\tt kekeli@guinness.ias.edu}
\bigskip
\centerline{Matthias Staudacher}
\smallskip
\centerline{Ecole Normale Sup\'erieure}
\centerline{\tt matthias@physique.ens.fr}
\bigskip
\noindent
We study two-dimensional gauge theories with fundamental fermions
and a general first order gauge-field Lagrangian.
For the case of $U(1)$ we show how standard bosonization of the
Schwinger model generalizes to give mesons interacting through a general
Landau-Ginzburg potential.
We then show how for  a subclass
of $SU(N)$ theories, 't Hooft's solution of large $N$ two-dimensional QCD can
be generalized in a consistent and natural manner.
We finally point out the possible relevance of studying these
theories to the string formulation of two-dimensional QCD as well as to
understanding QCD in higher dimensions.
\Date{January 13, 1994}
\nref\th{G. 't Hooft, Nucl. Phys. B75 (1974) 461.}
\nref\migdal{A.A. Migdal, Sov. Phys. - JETP 42, 413.}
\newsec{Introduction}
It is known that for two-dimensional Yang-Mills theory, the action $\tr
F^2$ does not enjoy the uniqueness of its higher dimensional counterpart.
In Migdal's lattice formulation \migdal,
the local Boltzmann weight for a plaquette $\D$ of area $a^2$ is
a heat kernel for the operator $\tr E^2$, given by a sum over
all irreducible representations of the gauge group:
\eqn\weight{Z_\D(U)=
\sum_R ({\rm dim} R)\,\e^{-g^2a^2\,{\cal C}(R)}\,\chi_R(U),}
where $U$ is the holonomy of the gauge field around the plaquette, $g$ is
the gauge coupling and ${\cal C}(R)$
is the second Casimir of the representation $R$.
Expanding about $U=1$ one shows that this approximates the continuum
Yang-Mills
action arbitrarily well as $a\rightarrow 0$, and
because of the well-known self-reproducing property, even with finite area
plaquettes the theory is exactly equivalent to a continuum theory.

The self-reproducing property is true for an arbitrary function ${\cal C}(R)$.
Such a lattice action \weight\ will correspond to a first-order continuum
action in $D=2$ of the form:
\eqn\first{S = \int d^Dx\, \big[ \tr (E\,\ep^\mn F_\mn) - g^2 f(E) \big] }
Here $F_\mn =\pa_\mu A_\nu -\pa_\nu A_\mu + i [A_\mu , A_\nu] $ is the
Yang-Mills field strength and $E$ is a scalar field in the adjoint
representation of the gauge group. Standard dimensional analysis applied
to \first\ gives $F_\mn$ dimension 2 and $E$ dimension $D-2$, so for $D=2$
power counting allows an arbitrary function $f$. As has been pointed out
\ref\witten{E. Witten, Comm. Math. Phys. 141 (1991) 153.},
the self-reproducing property of \weight\ or the area-preserving
diffeomorphism invariance of \first\ allows exact solutions for
arbitrary functions ${\cal C}$ and $f$, respectively.

In this paper, we study the generalized two-dimensional \su
gauge theory \first\ minimally coupled to fundamental fermions in the
large  $N$ limit. We demonstrate the consistency of this generalization
and present ample evidence that it is both highly non-trivial
and very natural. We thus add further examples to the
vast number of known universality-classes of two-dimensional field
theories and start addressing an
amusing mathematical question: How does this modification of the
gauge interaction affect the natural observables in QCD$_2$, namely the
spectrum of mesons and their $S$-matrix?
But the main motivations for our work are two-fold:
First, similar modifications of
the bare QCD$_4$ action might ultimately make this theory more tractable.
Second, it is extremely important to keep looking for exactly
solvable models whose analytical structure can be more fully understood or
which more closely resemble four-dimensional QCD.

Indeed, at present the
analytically tractable model that is most analogous to QCD$_4$ seems to be
QCD$_2$ in the large $N$ limit.\th~
To support this point of view, we first point out that no analytic
description exist of the spectrum of $D>2$ strongly coupled field
theories. The situation is better in two dimensions. For example, the
$D=2$ sigma models with $SU(N)$ target spaces have very
interesting analogies with QCD$_4$, notably logarithmic scaling, a mass
gap and dimensional transmutation,
however the spectrum and exact S-matrices show
fewer analogies, and there is no direct analog of confinement. Another
interesting possibility is to consider adding adjoint matter in QCD$_2$,
which gives a theory with roughly the number of degrees of freedom
expected in
a higher dimensional large $N$ gauge theory
\ref\kk{S. Dalley and I. R. Klebanov, Phys.~Lett.~B298 (1993) 79;\hfill\break
D. Kutasov, Chicago preprint EFI-93-30, Jun 1993. hep-th/9306013.}
and may in the near future be analytically tractable.
Finally, the finite $N$ model is perhaps not much harder to work with
numerically than the large $N$ theory; adopting light-cone gauge and
light-front quantization works equally magical simplifications at finite
$N$, the essential difference is the presence of quark loops, which have
been
verified to be unimportant for certain interesting quantities (e.g.\ low
lying masses).
\ref\who{See for example K. Hornbostel, Ohio State preprint
OHSTPY-HEP-T-92-020, Sep 1992;\hfill\break
K. Hornbostel, S. J. Brodsky, H. C. Pauli, Phys.~Rev.~D41 (1990) 3814;
\hfill\break
S. J. Brodsky and G. P. Lepage, in `Perturbative Quantum Chromodynamics,'
ed. A. H. Muller, World Scientific 1989, pp. 93-240 and references there.}
Despite this alternative, our own attitude at present is that we pay
little for the additional simplification of large $N$. Furthermore we find
the hopes for rewriting the theory as a string theory (which only makes
real sense for large $N$) attractive.

Pure gauge theory \first\ is exactly solvable for any $N$
and has no local degrees of
freedom. When fermions are introduced,  the non-quadratic nature of the
function $f(E)$ will introduce
complicated interactions and a general model will no longer be
solvable. Nevertheless, since the only dimensionful
coupling
constant in the theory, $g^2$, has positive dimension, one expects that
the
generalized theory remains super-renormalizable. As an example, we will
begin in section 2 by considering the simpler model of two-dimensional
$U(1)$
gauge
theory with fermions but with more general gauge interactions as in
\first.
This ``generalized Schwinger model''
can be studied by the standard bosonization method. While the original
(massless) Schwinger model effectively describes the meson
(fermion-anti-fermion bound state) as a free massive scalar, the
generalized
model is an interacting scalar theory with a general Landau-Ginzburg
potential.
In two dimensions, Laudau-Ginzburg scalar theory is super-renormalizable,
and
leads to a rich family of critical universality classes, described by
$c<1$
conformal field theories.

We continue our discussion in section 3 with a special class of
generalized QCD$_2$ in large $N$ limit, with gauge field action
\first\ and $f(E) = \sum_n f_n \tr  E^n$. With a single $\tr$ in \first,
we are able to follow 't Hooft's original solution and derive
the modified bound state equation as an explicit function of $f$.
The infrared problem already present in the original model turns
out to be even more subtle in our case and we have to carefully
evaluate a class of highly singular loop integrals. After dealing
with this technical problem we find that all infrared divergences
cancel out of the bound state equation, thus giving strong evidence
for the consistency of our models. The resulting modified equation
turns out to be mathematically very natural. So far we are
unable to present new exact solutions; however, we prove that
the massless state present for the standard theory in the chiral limit
(massless bare quarks) continues to be a solution for any function $f$,
as is expected on general grounds. After discussing qualitatively
how the linear force law is affected by the new terms in the action
we go on to present the results of a preliminary numerical study of
the simplest possible modification of the bound state equation,
produced by taking $f_3\ne 0$.
For small coupling the spectrum is a simple modification of that for the
original theory and perturbatively (in $1/N$) the theory appears sensible
and stable.
The lowest massive meson state (present for massive bare quarks) is driven
towards mass zero,
and a critical coupling $f_{3c}$ exists with a new massless state.
(This is in the free, leading order in $1/N$ theory -- more generally this
demonstrates the existence of critical points in these theories).
For larger coupling the spectrum contains a tachyon.

The potential relevance of our study to more physical dimensions
$D > 2$ is discussed in section 4.
If QCD$_2$ is studied as a simple analog of QCD$_4$, any of the
generalized theories introduced above has a priori an equal right to be
considered. More precisely, if an analytic technique exists which makes
sense
in  arbitrary dimension $2\le D\le 4$ and whose continuum limit is unique
in
higher dimensions, then the dimensional continuation of this unique theory
should give a preferred theory in $D=2$ which may or may not be described
by the $\tr F^2$ action.

Turning this around, if we find a member of the general class
of two-dimensional theories with some special simplicity,
perhaps this simplification has an analog in higher dimensions as well.
We discuss the possibility of using the generalized actions
in higher dimensions, which
generically produces a theory with the same continuum limit, but possibly
more convenient cutoff scale dynamics.

\nref\kostov{I. K. Kostov, SACLAY-SPHT-93-050, June 1993. hep-th/9306110.}
\nref\GT{D. J. Gross and W. Taylor, Nucl. Phys. B400 (1993) 181 and
 Nucl. Phys. B403 (1993) 395.}
In particular we propose that to learn more about the connection with
string
theory it may be necessary to consider the general gauge theories, since
a simple string theory may give a complicated spacetime theory.
We will argue that the string picture that emerges from recent
work on large $N$ QCD$_2$
\refs{\GT,\kostov}
can be simplified considerably if one allows generalized gauge
interactions in the target spacetime.

Section 5 contains concluding remarks.

\newsec{The generalized Schwinger model.}

The Schwinger model describes standard $U(1)$ gauge theory minimally
coupled to
Dirac fermions in two space-time dimensions. While being exactly solvable
(for
the massless case), it demonstrates a rich spectrum of phenomena, such as
the vacuum angle, quark confinement and the Higgs mechanism, and has been
studied
as a toy model for more physical theories in four dimensions. Detailed
accounts
of
the model and its properties have been given in many different approaches
\ref\schwinger{See, e.g., S. Coleman, R. Jackiw, and L. Susskind, Ann. Phys. 93
(1975) 267;\hfill\break S. Coleman, Ann. Phys. 101 (1976) 239.}
and will not be reviewed here. We will instead consider its
generalization in the sense of \first, and point out some new features.

The Lagrangian for the generalized Schwinger model is
\eqn\sch{L= {1\over\pi}E\,\ep^\mn F_\mn - g^2 f(E)
+ \overline\psi\gamma^\mu (i\pa_\mu - A_\mu) \psi - m\overline\psi\psi,}
In this first-order formalism, the usual Maxwell theory corresponds to
taking the scalar potential $f(E)=2 E^2/\pi^2$.
A linear term in $f$ would be the usual theta term.
The field $E$ is a pseudoscalar and thus if $f(E)$ is not an even
function,
parity will be explicitly broken in the model.

The field-theoretic aspects are most clearly exhibited by
bosonization (originally discovered in this context). The standard rules
are:

\eqn\bosoniz{\eqalign{
\overline\psi\gamma^\mu\psi = \emn\pa_\nu\phi/\pi\,,\cr
\overline\psi i\gamma^\mu\pa_\mu\psi =
{1\over 2\pi}\pa^\mu\phi\pa_\mu\phi\,,\cr
\overline\psi\psi =cm\cos(2\phi)\,,\cr}}
where $c$ is a constant related to the normal-ordering in defining the
composite operator.

In the second-order formalism of the usual Schwinger model, one chooses a
convenient gauge (such as axial gauge) so that the gauge field equation
contains no time derivative, and can be solved as a constraint equation.
The same argument applies to the first-order formalism and a general
potential $f(E)$. The bosonized Lagrangian can be written as

\eqn\schh{L= {1 \over\pi}\emn\pa_\mu A_\nu\ (E-\phi) - g^2 f(E)
+ {1\over 2\pi}\pa^\mu\phi\pa_\mu\phi - cm^2\cos(2\phi)\ .}
The gauge potential appears linearly and serves as a Lagrangian
multiplier.
Fixing a gauge $A_1=0$ and integrating out $A_0$ simply gives a constraint
that
determines the scalar field $E$ in terms of $\phi$:
\eqn\cst{ E= \phi + {\theta\over 2}}
where the constant $\theta$ is the vacuum angle of the theory.
After a convenient shift of $\phi$ by $\theta/2$, the final
Lagrangian is:
\eqn\final{L={1\over 2\pi}\pa^\mu\phi\pa_\mu\phi - g^2 f(\phi)
- cm^2\cos(2\phi - \theta)\ .}
In the massless case ($m=0$), the generalized Schwinger model simply
describes
an interacting scalar (the meson) with a general Laudau-Ginzburg potential
$f$.

{}From this example, we see that a generalized two-dimensional gauge
theory
coupled with fermions remains super-renormalizable for a general potential
$f$,
as suggested by the positive dimension of the gauge coupling.
The naive intuition that higher powers of $F$ would be irrelevant is
incorrect.
The essential reason is that the gauge field has no local degrees of
freedom.
Power counting should be done with dimensions set by the fermion
Lagrangian,
giving $A$ dimension $1$ and $E$ dimension $0$.
Unlike pure $D=2$ Yang-Mills, we must give a renormalization prescription
to completely define the model and determine the mapping between bare
Lagrangians and physical models.  A simple and standard choice would be to
normal order the interaction using the free mass $m$ bosonic propagator
for
contractions.

Mesons exist as bound states of fermions, but they may have more
complicated
interactions determined by $f$.  The qualitative physics of the
generalized
model is typically similar to the original model, but a number of
interesting
modifications are possible.

First, there is the possibility of multiple vacua or even instability
of the theory.
The function $f(E)$ gives the energy per
unit length of a pure gauge field configuration of strength $E$.
After bosonization, the function $V(E)=f(E) + m^2 \cos (2E - \theta)$ has
become a potential, and each minimum of $V(E)$ is a possible vacuum of the
theory.

If we have multiple
minima $E_i$, we have the possibility of stable states of non-zero charge
$q=n\pi$
such that both left and right asymptotic fields are minima, if $q=E_i-E_j$
for some $i$ and $j$.
The choice $f(E)=\cos 2 E$ gives a particularly simple illustration.
Each
fermion now becomes a soliton with a gauge field `dressing,' which
modifies
its mass while maintaining its non-interacting nature.

Second, we can tune to critical points of the interacting scalar theory.
The simplest critical point is attained by tuning the renormalized mass to
zero
in a theory with generic higher order interactions.  Perturbatively this
would
even have been possible in the original massive Schwinger model by taking
$g^2$
negative; however such a theory would have energy unbounded below
non-perturbatively.  In the generalized model we can add a higher power of
$E$
with positive coefficient to fix this problem.
The resulting theory has a phase transition; for $g^2 < g_c^2$ the vacuum
breaks the $Z_2$ charge conjugation symmetry $\phi\rightarrow -\phi$.
Near $g^2 = g_c^2$ we see the critical behavior of the Ising model.

The generalized Schwinger model should be a good qualitative quide to
phenomena we can expect in generalized QCD$_2$.
A point to keep in mind however is that our large $N$ treatment will only
discuss the single meson sector.
Clearly some of the field theoretic phenomena we described
above will already have signals in this sector.
For example, a critical point will be signaled by a meson mass
going through zero as a function of the couplings.
However, we will hardly be able to make a complete analysis
just knowing the spectrum.
The proper tool to study these phenomena would be the effective field
theory for the mesons.

Instabilities of the theory may or may not be visible in this sector.
The reinterpretation of $f(E)$ as a potential followed once the constant
mode
of $\phi$ became dynamical, which is very much a field-theoretic effect.
If we are constrained to the single meson sector, and have boundary
conditions
$E=E_0$ for the the gauge field,
we should only expect to see instabilities which can be detected by
considering
gauge fields with $|E-E_0|\le q$.
This is a classical argument of course and the actual situation will be
somewhat more complicated, as we will see.

\newsec{The generalized 't Hooft model.}

The Lagrangian is

\eqn\lag{
\CL = {N\over 8\pi}~ \tr E~ \epsilon^{\mu\nu}F_\mn -
{N\over 4\pi}~ g^2 \sum_{n=2}^{\infty} f_n ~\tr E^n
+\bar \psi (i \gamma^\mu D_\mu - m) \psi,}
where $E$ and the gauge potential $A_\mu$ are $N\times N$ hermitian
matrices.
The field strength is $F_\mn=\partial_\mu A_\nu - \partial_\nu A_\mu +
i [A_\mu,A_\nu]$ and the covariant derivative is $D_\mu=\partial_\mu +i
A_\mu$.
The $N$'s have been introduced to make planar diagrams survive
in the limit $N\rightarrow\infty$, with $f_n$ and $g$ fixed.  The $f_n$'s
are dimensionless and $g$ has dimensions of mass.
We take one flavor of fermionic quark in the fundamental.
We have not taken the most general function $f(E)$ of the
introduction; powers of traces are also possible
but will not be treated in this paper.
If we take $f_2=1/8\pi$, all other $f_n=0$
and integrate out $E$ we obtain the model of \th.

As in \th, taking light-cone gauge $A_{-}=0$ simplifies the theory
tremendously.  The preliminary observation that we eliminate the gauge
field self-interaction is vital but this would also have been true in any
axial or temporal gauge $n \cdot A=0$.
What makes light-cone gauge simplest (and the gauge in which the theory
has been solved most completely) is that we can then do light-front
quantization of a theory with an instantaneous (in $x^{+}$) interaction.
In light-front quantization virtual pair creation is impossible, because
of conservation of $p_{-}$ and the positivity of $p_{-}$ for every degree
of freedom (assuming $m>0$).
Physical quark pair creation is subleading in $1/N$.
Thus Hamiltonian evolution preserves the $\psi\bar \psi$ subsector of the
Fock
space and we can completely integrate out the gauge field -- the quark
self-energy is entirely reproduced by normal ordering.

Although in some ways the light-front Hamiltonian description is more
physical,
the reduction to planar diagrams is clearer in Lagrangian perturbation
theory, as used in \th, so our derivation of the bound-state
integral equation will start there.

In light-cone gauge, the Lagrangian reduces to
\footnote*{Let $\gamma^0=\sigma_1$, $\gamma^1=i\sigma_2$,
$x^{+}=(x^0+x^1)/\sqrt{2}$, $\psi^t = (\psi_L,\psi_R)$, then $g^{+-}=1$,
$\epsilon^{+-}=-1$; and absorb a factor of $\sqrt{2}i$ into $\bar\psi$.}
\eqn\llca{\eqalign{
&{N\over 4\pi}~ \tr E\pam A_{+} -
{N\over 4\pi} ~g^2 \sum_{n=2}^{\infty} f_n \tr E^n\hfill\cr+
&\psi^{+}_L\pam \psi_L + \psi^{+}_R(\pap+iA_{+}) \psi_R
- {m\over\sqrt{2}}(\psi^{+}_L \psi_R + \psi^{+}_R \psi_L)}}
and after solving for $\psi_L$,
\eqn\llc{{N\over 4\pi}~ \tr E\pam A_{+} -
{N\over 4\pi} ~g^2 \sum_{n=2}^{\infty} f_n \tr E^n
+ \psi^{+}_R(\pap+{m^2\over2\pam}+iA_{+}) \psi_R .}

We see that there are $\la E A\ra$ propagators,
but no $\la E E\ra$ or $\la A A\ra$ propagator
(we can regard $\Tr E^2$ as a vertex).
Thus each propagator from an $E^n$ vertex
must be connected to a quark line.  This is what makes the model easily
solvable -- the gauge field self-interaction does not produce arbitrary
fishnet diagrams but only a simple generalization of the rainbow and
ladder diagrams of \th.
The Feynman rules are given in fig. 1, and a representative planar diagram
is in fig. 2.
\fig{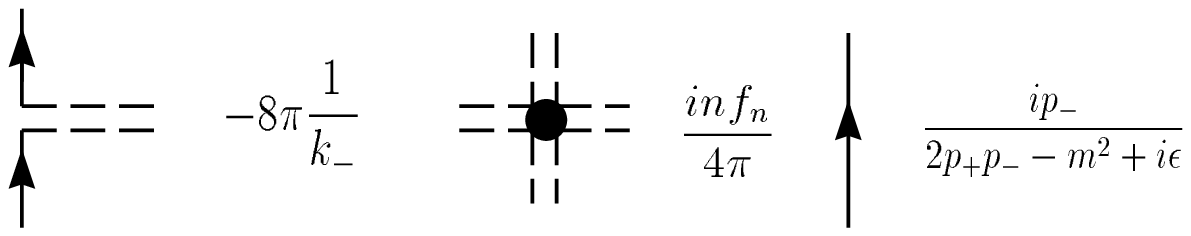}{Fig. 1.  Planar Feynman rules in the light-cone gauge.}

We will discuss the $q\bar q$ bound state.
(It is easy to see that the generalized interactions do not change
the fact that exotic mesons are not bound in leading order in $1/N$.)
As for 't~Hooft, the problem splits into two steps; evaluating the
renormalized quark propagator, and then the renormalized Bethe-Salpeter
kernel.
Planar diagrams are generated recursively as in \th, and the new
elements in the quark self-energy are the ``M'' and higher order diagrams,
while the $O(f_n)$ correction to the kernel is a sum of graphs with $l$
and $n-l$ gauge legs attached to the quark and antiquark.
All of these graphs can be expressed in terms of the ``master integral''
\fig{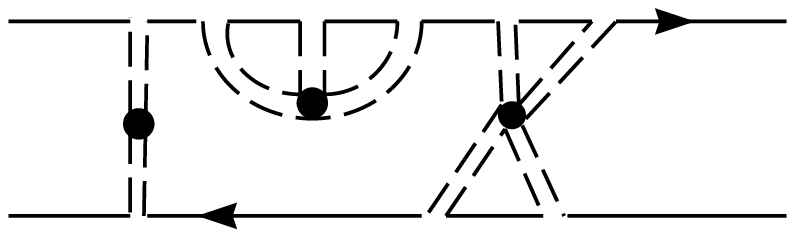}{Fig. 2.  A representative planar diagram.}
\eqn\master{\eqalign{I_n(p,p') = &({2\over \pi})^{n-1}
\int_{-\infty}^{\infty} d^2k_1 \ldots d^2k_{n-1} \cr
&{1\over p_{-}-k_{1-}}
S(k_1) {1\over k_{1-}-k_{2-}} S(k_2) \ldots \tcbreak{\qquad}
{1\over k_{n-2,-}-k_{n-1,-}} S(k_{n-1}) {1\over k_{n-1,-}-p'_{-}}.}}
with $n\geq2$ where $S(p)$ is the renormalized quark propagator
\eqn\prop{S(p)={i p_{-} \over 2 p_{+} p_{-} -m^2- p_{-} \Gamma(p_{-}) +i
\epsilon}}
The self-energy is
\eqn\self{-i \Gamma(p) = -2 g^2 \sum_{n=2}^{\infty} (-i)^{n-1} n f_n
I_n(p,p)}
and the 2PI kernel is
\eqn\kernel{K(q,q'; p,p') = 4 g^2 \sum_{n=2}^{\infty} (-i)^n n f_n
\sum_{l=1}^{n-1}
I_l(q,q') I_{n-l}(p'-q',p-q)}
where $I_1(p,p')=1/(p-p')$, and the incoming and outgoing quark (resp.
antiquark) momenta are $q$ and $q'$ (resp. $p-q$ and $p'-q'$).

As in \th, it is important that the self-energy does not depend on $p_{+}$
(as is clear from \master).
Physically this is because the interaction is instantaneous in $x^+$.
This allows us to do the $k_{i,+}$ integrals independently without
knowing $\Gamma$ or $m$, producing
\eqn\mastertwo{\eqalign{I_n(p,p') = &
\int_{-\infty}^{\infty} dk_1 \ldots dk_{n-1} \cr
&{1\over p-k_{1}}
\sgn k_1 {1\over k_1-k_2} \sgn k_2 \ldots
{1\over k_{n-2}-k_{n-1}} \sgn k_{n-1} {1\over k_{n-1}-p'}.}}
We have dropped the light cone index $\{-\}$ in \mastertwo.
This result could also be derived by normal ordering the terms of
quadratic and higher order in the LC Hamiltonian.  The $\sgn k_-$ there
comes from fermi statistics and the expansion of the field in modes with
$k_->0$.

The $k_{-}$ integrals are infrared divergent and must be regulated.
The simplest prescription would be to follow
\ref\ccge{
C. G. Callan Jr., N.Coote, and D. J. Gross, Phys. Rev. D13 (1976) 1649.}
and define the $k_{-}$ integrals as principal part integrals.
However we will immediately run up against a subtlety first pointed out in
this context by T. T. Wu
\ref\wu{T. T. Wu, Phys. Rep. 49 (1979) 245.}:
principal part integrals do not commute in general.
A physical and safe procedure is to regulate the kernels ${1 \over p-p'}$
by cutting them off at some small momentum $\lambda$, as 't Hooft did
originally \th. It is important to use the {\it same} $\lambda$ for
all kernels in \mastertwo\ and safest to take $\lambda$ to zero only after
performing
all integrations.\footnote{*}{
It is by now clear that if properly derived from a physical regulator
(which
can be implemented in the original action),
the principal part prescription, as used originally in
\refs{\th,\ccge}\ and as will  appear
in our intermediate results below, is justified.  One way to distinguish
the
situations in which it is valid is to note that in a Hilbert transform
acting
on continous functions of an appropriate type, orders of integration can
be
exchanged.  In particular, the p.p. integral in our bound-state equation
will
act on smooth wave functions.  A completely independent check of the
spectrum found in \th\ has also been done in
\ref\lwb{M. Li, L. Wilets and M. C. Birse, J. Phys. G 13 (1987) 915.},
using the axial gauge treatment of
\ref\bars{I. Bars and M. B. Green, Phys. Rev. D17 (1978) 537.}.}
A naive treatment of singularities by principal part
integration in this problem leads to wrong results;
i.e.~they differ from the $\lambda\rightarrow0$ limit
of the finite $\lambda$ result.
We found it convenient to invoke a slightly smoother
version of 't Hooft's sharp cutoff; i.e. we regulate
\eqn\lker{{1 \over p-p'} \rightarrow \half \bigg( {1 \over p-p'+i \lambda}
+{1 \over p-p'-i \lambda} \bigg) }
It is then straightforward to explicitly work out the first of the above
regulated integrals:
\eqn\itwolam{\eqalign{I^{\lambda}_2(p,p')=&\half \bigg( {1 \over p-p'} +
{p-p' \over (p-p')^2 + 4 \lambda^2} \bigg)
\log {{p^2+\lambda^2} \over {p'^2+\lambda^2} } \cr
&-{2 \lambda \over (p-p')^2 + 4 \lambda^2} \big( \arctan {p \over \lambda}
+\arctan {p' \over \lambda} \big) }}
Taking now $\lambda$ to zero one finds
\eqn\itwo{I_2(p,p')={2 \over p-p'} \log \big \vert {p \over p'} \big\vert
- \pi^2 \sgn p \; \delta(p-p')}
It is not possible to express the regulated $I^{\lambda}_n$'s with $n>2$
in terms of elementary functions.
We must therefore find an alternative method to deduce the correct
limit of these multiple singular integrals.
It is convenient to introduce the generating function
\eqn\itilde{u(p,p'; z) = \sum_{n=0}^{\infty} z^{-n} I_n(p,p')}
and define
\eqn\istart{\eqalign{I_0(p,p') =& \sgn p \; \delta(p-p'), \cr
I_1(p,p') =& {{\cal P} \over p-p'} } }
where ${\cal P}$ denotes the Cauchy principal value.
Formally, the generating function $u(p,p'; z)$ satisfies the integral equation
\eqn\ueq{z~u(p,p';z)=z~\sgn p \; \delta(p-p') +
\int_{-\infty}^{\infty} dk~{{\cal P} \over p-k}~~\sgn
k~~u(k,p';z)}
We can give it meaning by noting that $\sgn p~\int dk {{\cal P} \over p-k}$
should be a well defined integral operator on continuous functions; in
particular
the functions $\sgn k~\vert k \vert^{2 \nu}$ are eigenfunctions of this
operator:
\eqn\eigen{\sgn p~\int_{-\infty}^{\infty} dk {{\cal P} \over p-k}
\sgn k~\vert k \vert^{2 \nu}=\pi~\cot \pi \nu~\sgn p~\vert p \vert^{2
\nu}}
In view  of eqs.\mastertwo, \itilde\ these functions are also
eigenfunctions
of the integral operator $\sgn p~\int dk~u(p,k;z)$:
\eqn\ueigen{\sgn p~\int dk~u(p,k;z)~\sgn k~\vert k \vert^{2 \nu}=
{z \over z-\pi \cot \pi \nu}~\sgn p~\vert p \vert^{2 \nu}}
We therefore have to find a kernel which reproduces \ueigen. This is
done by noticing that $\sgn p~\int dk {{\cal P} \over p-k}~
\vert {p \over k} \vert^{2 \alpha}~\sgn k~\vert k \vert^{2 \nu}=
\pi \cot \pi (\nu - \alpha)~\sgn p~\vert p \vert^{2 \nu}$. Using the
addition formula for $\cot \pi (\nu - \alpha)$ one finds the (unique)
expression
\eqn\usoltn{u(p,p';z)={z \over z^2+\pi^2} \big[
{{\cal P} \over p-p'}~\vert {p \over p'} \vert^{2 \alpha(z)}+
z~\sgn p~\delta(p-p') \big]}
where $z=\pi \cot \pi \alpha(z)$, and thus $2 \alpha(z)=
{2 \over \pi} \arctan {\pi \over z}$.
One can check by direct computation, using the formula (derived from
\lker)
\eqn\kerker{{{\cal P} \over p-k}~{{\cal P} \over k-p'}=
{{\cal P} \over p-p'}~\big({{\cal P} \over p-k}+{{\cal P} \over k-p'}
\big)-
\pi^2~\delta(p-p')~\half~\big(\delta(p-k)+\delta(k-p') \big)}
that $u(p,p';z)$ is indeed the solution of the integral equation \ueq.
Expanding in $z^{-1}$ to ${\cal O}(z^{-4})$ one finds in addition to
\istart,\itwo
\eqn\inext{\eqalign{I_3(p,p')=&2 {{\cal P} \over p-p'} \log^2 \vert{p
\over p'}
\vert
-\pi^2{{\cal P} \over p-p'} \cr
I_4(p,p')=&{4  \over 3} {{\cal P} \over p-p'} \log^3 \vert{p \over p'}
\vert-
{8 \over 3} \pi^2 {{\cal P} \over p-p'} \log \vert {p \over p'}\vert +
\pi^4
\sgn p~\delta (p-p')}}
It is important to notice that these loop integrals are generalized
functions (distributions). One easily verifies to low order using
these explicit formulae together with \kerker\ that we can recursively
calculate the higher $I_n$'s from lower orders and that
the result does not depend on the order of integration.
In order to test the independence of our results on the details of the
infrared
cutoff procedure we numerically checked our formulae to this order using
't~Hooft's
sharp cutoff. We found convergence to the above result \inext\ as the
sharp cutoff tends to zero.

In order to calculate the contributions to the self-energy $-i\Gamma(p)$
we
take the limit $p\rightarrow p'$ of the distributions $I_n(p,p')$.
All terms with powers of $\log \vert {p \over p'} \vert$ higher than one
tend to zero while the terms linear in the logarithm give a finite,
non-zero
`mass renormalization'.%
\footnote*{Strictly speaking, this is not a mass renormalization, as is
clear
from considering a zero bare mass quark, which would be protected from
perturbative mass
renormalization by chiral symmetry.  The choice of LC gauge breaks Lorentz
invariance in gauge-variant quantities and allows producing a self-energy
which
enters the same way as a mass in the subsequent analysis.  We thank N.
Seiberg
for a discussion on this point.}
The terms ${{\cal P} \over p-p'}$ tend to zero likewise
due to our prescription \lker: ${{\cal P} \over p-p'}\vert_{p=p'}=0$.
We conclude from \usoltn\ that the odd order
mass renormalizations are zero. The even order self-energies are both
finitely
and infinitely renormalized; the latter due to the presence of terms
$\sim \delta(0)$. Such infinite self-energies are already present in the
original 't~Hooft model and we have to study the full bound-state equation
before getting worried. Following again \th\ it is given by
\eqn\prebs{\big[ 2 p_+ -m^2 \big({1  \over q}+{1 \over p-q} \big) \big]
\phi(q)=\big[\Gamma(q)+\Gamma(p-q)\big] \phi(q) +
\int_{0}^{p} dk~K(q,k;p,p)~\phi(k) }
All momenta without index are to be understood as minus-components. We can
express this equation through the generating
function $u(p,p';z)$; this leads for the right-hand side to the result
\eqn\nextbs{\eqalign{2 g^2 \oint {dz \over 2 \pi i} f'(z)
\bigg[ -\big( u(q,q;i z) & +u(p-q,p-q;i z) \big)~\phi(q)+ \cr
& +2 \int_{0}^{p} dk~u(q,k;i z)~u(p-k,p-q;i z)~\phi(k) \bigg] }}
where $f(z)$ is as in \lag. Using the identity
\eqn\kersqr{\big( {{\cal P} \over q-k} \big)^2=
{{\cal P} \over (q-k)^2}+\pi^2~\delta(0)~\delta(q-k) }
which is again derived from \lker\ we can rewrite after some algebra
the expression \nextbs:
\eqn\explbs{\eqalign{2 g^2 \oint {dz \over 2 \pi i} f'(z)
&\bigg[ -i z{z^2+\pi^2 \over (z^2-\pi^2)^2} 2 \alpha(i z)
\big( {1 \over q}+{1 \over p-q} \big)~\phi(q) \cr
&\qquad -{2 z^2 \over (z^2-\pi^2)^2} \int_{0}^{p} dk {{\cal P} \over
(q-k)^2}
\big(  {q(p-k) \over (p-q)k} \big)^{2 \alpha(i z)}~\phi(k)\bigg] } }
All the infrared divergences, i.e.~all terms with $\delta(0)$ have
canceled!
This is the generalization of the correponding phenomenon in the
ordinary 't Hooft model.  It serves as a highly nontrivial check on
the internal consistency of our models. Note also that a further
mass renormalization coming from the kernel is seen in \explbs;
it remains however true that only the even couplings in the action
lead to mass renormalization. After going to dimensionless variables
in the usual way, $\gamma=\pi {m^2 \over g^2}$,
$\mu^2={\pi \over g^2} 2 p_{+} p_{-}$ and $x={q \over p}$, $y={k \over p}$
one obtains the final form of the bound-state equation:
\eqn\bs{\eqalign{
\big[ \mu^2 - \gamma \big({1 \over x} + & {1 \over 1-x} \big)
\big]~\phi(x)=
\cr
&= 2 \pi \oint {dz \over 2 \pi i} f'(z)
\bigg\{ -i z{z^2+\pi^2 \over (z^2-\pi^2)^2}~2 \alpha(i z)~
\big( {1 \over x}+{1 \over 1-x} \big)~\phi(x) \cr
& \;\;\;\;\;\;
 -{2 z^2 \over (z^2-\pi^2)^2} \int_{0}^{1} dy {{\cal P} \over
(x-y)^2}
\bigg[  {x(1-y) \over (1-x)y} \bigg]^{2 \alpha(i z)}~\phi(y) \bigg\} } }
Eq. \bs\ constitutes the principal technical result of the present work.
The generalized kernel results in logarithmic corrections to the
't Hooft equation; for example, turning on the couplings $f_3$ and $f_4$
in
addition to $f_2=1/8\pi$ in the 't Hooft model one obtains
\eqn\bsexp{\eqalign{\mu^2 \phi(x)  =& \
(\gamma -1 -{160 \over 3}\pi^3~f_4)~\big( {1 \over x}+{1 \over 1-x}
\big)~\phi(x) \cr
& -\int_{0}^{1} dy {{\cal P} \over (x-y)^2} ~\big[
{}~(1+32 \pi^3~f_4)  -  24 \pi ~i~f_3~\log {x(1-y) \over (1-x)y} \cr
& \hskip2.0in\relax
-32\pi~f_4~\big( \log {x(1-y) \over (1-x)y}\big)^2 \big]~\phi(y) } }
Note the factor of $i$ in front of the term linear in the logarithm,
which came from the $k_+$ integral producing \master.
It serves to ensure the hermiticity of our Hamiltonian.
Indeed, as is easily seen from \bs,
all terms with odd powers of the logarithm (corresponding to odd
potentials)
come with such a factor of $i$, while even powers are real.

To complete the discussion we must find boundary conditions for $\phi(x)$
compatible with self-adjointness of the Hamiltonian implied by \bs\ . As
in the 't~Hooft model, the appropriate boundary condition is
$\phi(x)=x^\beta$ at
$x=0$ and $\phi(x)=(1-x)^\beta$ at $x=1$, where the exponent $\beta$ is
determined by requiring the leading term $x^{\beta-1}$ in the integral
equation \bs\ to cancel. It must satisfy
\eqn\bd{\eqalign{
\gamma = 2 \pi \oint {dz \over 2 \pi i} f'(z)
\bigg[& i z{z^2+\pi^2 \over (z^2-\pi^2)^2}~2 \alpha(i z) \tcbreak{}
  - {2 z^2 \over (z^2-\pi^2)^2} \big( \pi\beta-2\pi\alpha(iz) \big)
\cot \big( \pi\beta-2\pi\alpha(iz) \big) \bigg] }}

Let us discuss some of the properties of the bound-state equation. A very
useful identity is
\eqn\bswave{\int_{0}^{1} dy {{\cal P} \over (x-y)^2}~
\big( {1-y \over y} \big)^{2 \nu}=
-2\pi \nu~\cot 2 \pi \nu~\big({1 \over x}+{1 \over 1-x})~
\big({1-x \over x} \big)^{2 \nu} }
It may be used to show with a little bit of algebra
that for massless quarks ($m^2=0$, i.e.\ $\gamma =0$)
the wavefunction $\phi(x)=1$ solves the bound-state equation with
mass eigenvalue $\mu^2=0$ for an {\it arbitrary potential} $f(z)$.
We conclude that the massless ground-state is invariant under the
generalized interactions! While being true in general, we can make this
property manifest for a purely even potential by rewriting \bs\
for that subclass of models as
\eqn\evenbs{\eqalign{
\big[ &\mu^2 - \gamma \big({1 \over x} +  {1 \over 1-x} \big)
\big]~\phi(x)=
\cr
& - 2 \pi \oint {dz \over 2 \pi i} f'(z)
{2 z^2 \over (z^2-\pi^2)^2} \int_{0}^{1} dy {{\cal P} \over (x-y)^2}
\bigg[  {x(1-y) \over (1-x)y} \bigg]^{2 \alpha(i z)}~
\big(\phi(y) - \phi(x) \big) } }
%

The existence of such a massless state in the 't Hooft model is a
consequence
of chiral $U(1)$ symmetry as shown in
\ref\afflect{I. Affleck, Nucl. Phys. B265 [FS15] (1986) 448.}.
Although in \lag\ both $SU(N)$ and $U(1)$ vector symmetry are gauged,
the large $N$ limit takes the $U(1)$ gauge coupling to zero in a way
that suppresses the chiral $U(1)$ anomaly.
The simplest argument for this is that the anomaly receives contributions
from
Feynman diagrams with one fermion loop, which by standard large $N$
counting is $O(1/N)$. This remains true in the generalized models.
Conservation of the vector $U(1)$ current in two dimensions allows writing
$J^\mu = \epsilon^{\mu\nu}\partial_\nu\phi$, and
conservation of the chiral $U(1)$ current $J_\mu^5=\epsilon_{\mu\nu}J^\nu$
then implies the existence of a free massless boson
\ref\witt{E. Witten, Nucl. Phys. B145 (1978) 110.}.

A more detailed discussion using non-abelian bosonization is given in
\afflect.
The fermions are represented by a $k=1$ $SU(N)$ WZW model
with a massless boson representing the remaining $U(1)$. The complete
theory
consisted of a minimally gauged $SU(N)$ WZW model and a massless boson
$\phi$
coupled with the $U(1)$ gauge field $\tr F$ through $\phi \tr F$.
For the 't Hooft model there is no coupling between $SU(N)$ and $U(1)$
sectors.
In the generalized models, terms $\tr E^n$ with $n\ge 3$ produce
explicit couplings between the $SU(N)$ and $U(1)$ gauge fields
at subleading order in $1/N$.  This surely implies that at
finite $N$ long-distance physics eliminates the massless boson, and
the bosonized action
would be a good starting point for analyzing this.

We have found that adding $\tr E^n$ to the action adds new terms to the
integral kernel in the bound-state equation up to
${1\over(p-p')^2}\log^{n-2}(p/p')$.
This is essentially the Fourier transform of the interquark potential.
Only the even terms $\tr E^{2n}$ change the strength of the
linear confining potential,
while odd terms $\tr E^{2n+1}$ produce short range forces.
The long-distance behavior of the potential can be understood classically.
Solving Gauss' law $\partial_- E = {1\over N}J_-$ for a source at $x^-=0$
produces $E = q \epsilon(x^-)$ (with $\epsilon(x)=\theta(x)-1/2$).
Solving the equation of motion $\partial_- A_+ = f'(E)$
(with boundary conditions such that the background potential is zero)
then reproduces qualitatively the behavior we found.
However, the precise strength of the induced linear potential is not given
correctly by such a simple argument for the new interactions $n\ge 4$.
One reason for this is that it does not reproduce the planar nature of the
large $N$ perturbation theory.

The new terms in the bound-state kernel are less singular at $k=k'$ and
correspond to short range forces.  Clearly they can only be understood in
the quantum theory.  The terms
${1\over(x-y)^2}\log^{n-2}(x(1-y)/y(1-x))$ with $n>2$
in the bound-state kernel are analytic at $x-y=0$ (zero momentum transfer)
for $x$ and $y$ away from the endpoints $0$ and $1$.
If we neglected the endpoint singularities,
their Fourier transforms to position space would therefore be contact
terms.
In the limit of heavy quark masses this would be justified; the
wavefunction
would be highly peaked (for equal quark masses) at $x=1/2$.
The naive four-quark operators corresponding to these contact terms
would be
non-renormalizable even in two dimensions, but in the sense of an
effective
theory they would give a good description.
Away from the heavy quark mass limit, the endpoint singularities become
important, and it is not clear to us whether a local effective field
theory
description exists.
Related to this,
since the kernel is not only a function of momentum transfer,
the interactions also have the
slightly strange feature that the positions $x^-$ of the quarks can change
instantaneously in $x^+$ during the interaction.

The new interactions will shift the masses of all mesons except the
massless
one (present for $m_q=0$).
Using perturbation theory to estimate this, we find that for an even
perturbation $\tr E^{2n}$ there is a
correction at first order, with the same sign as the coupling, while for
an odd perturbation $\tr E^{2n+1}$ the first correction is at second order
and always negative.
A WKB argument shows that for the highly excited states
$m_n^2 \sim n$ the $\log^k$ coupling shifts $m_n^2$ by $O(n^{1-k})$.
We therefore expect that for sufficiently strong coupling (negative for
even
perturbations) we can drive a meson massless or even tachyonic.
Classically one certainly expects instability for certain potentials,
as we mentioned earlier.

\nref\num{A. J. Hanson, R. D. Peccei and M. K. Prasad, Nucl.~Phys.~B121
(1977) 477;\hfill\break
R. L. Jaffe and P. F. Mende, Nucl.~Phys.~B369 (1992) 189.}
We have preliminary numerical results for the case $f_3\ne 0$,
at the special mass $\gamma=1$.
Our numerical method is based on the one developed in Hanson et. al. \num~
and is described in the appendix.
For the 't Hooft model with coupling $f_2=1/8\pi^3$ (chosen to eliminate
some
$\pi$'s from the following) and $\gamma=1$, the spectrum is (combining
known
analytic and numerical information)
$m_n^2 \sim n - 0.25 - 0.01/n^2 + \ldots$~.

\fig{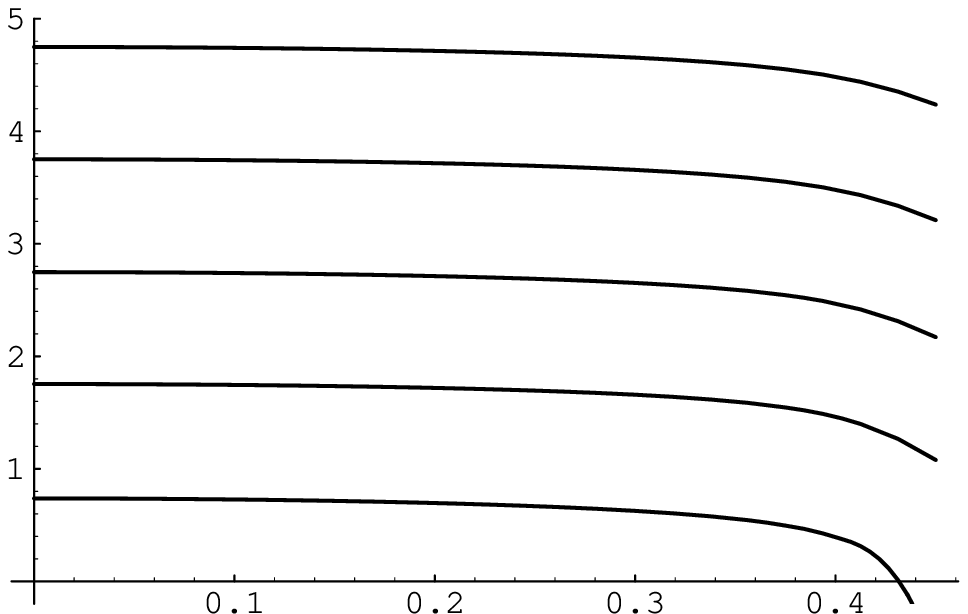}{Figure 3.  The first five mass levels as a function of
$\epsilon$.}
\fig{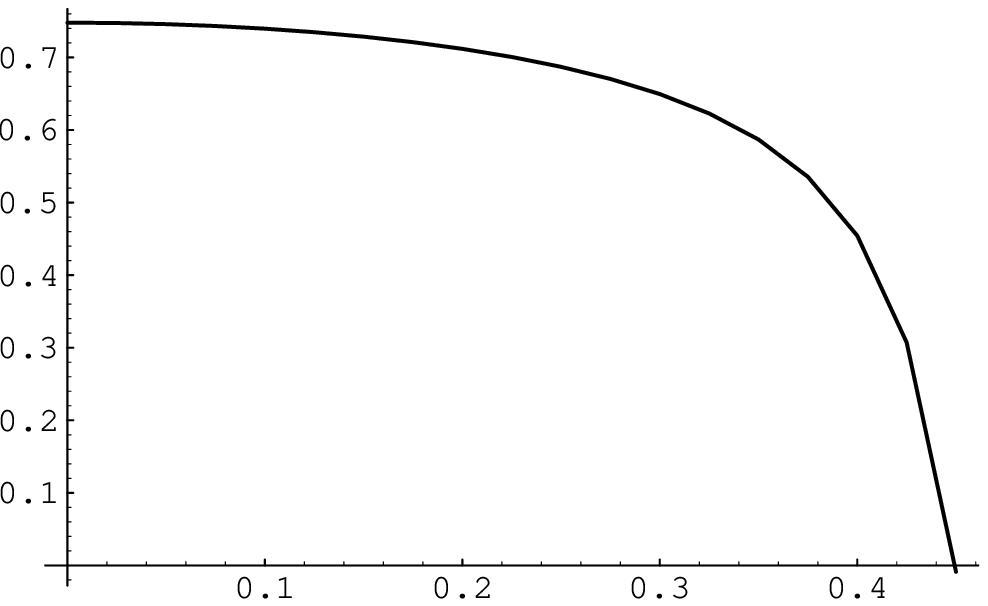}{Figure 4.  Fit to the intercept as a function of $\epsilon$.}

Our results for $f_3 \equiv \epsilon/24\pi^3 \ne 0$
are consistent (at the $1\%$ level
for $\epsilon<0.3$) with the slope remaining unchanged,
and the
intercept $0.75$ decreasing to $0.75-1.04 \epsilon^2$
for small $\epsilon$, consistent with second-order perturbation theory.
There is a critical coupling $\epsilon_c \sim 0.45$ where the lightest meson
goes
through zero mass.  We interpret this as the instability we discussed in
section 2.  For $|\epsilon|<\epsilon_c$ we see no sign in this calculation of
instability, though one might expect it to show up in a multi-meson
sector.
The main point is simply that the model in this sector is well-defined and
different from the original 't Hooft model (for any quark mass).

We expect models with $f_{2n}>0$, $f_{2n+1}=0$ to be completely
well-defined
and stable.
More generally, we expect that $f(E)$ bounded below suffices, if we derive
our
bound state equation by expanding about the true minimum.

For very large $f_3$, it is not clear whether the model will confine.
Standard techniques for singular integral equations
\ref\musk{N.I. Muskhelishvili, Singular Integral Equations, Dover.}
should suffice to solve the pure $f_3$ model (since the kernel has a
single
pole) and one expects the integral operator in this case to have
continuous spectrum.
Very naively, a small $f_2$
coupling looks like a singular perturbation which could restore
the discrete spectrum and possibly confinement.
However, it could be that the correct boundary condition here is
given by a different branch of \bd\ than one takes for small $f_3$,
which might still allow continous spectrum (as one expects from
the large mass limit.)

As one approaches the critical value $\epsilon_c$, the end-point behavior of
the wave functions appears to approach $x^\beta$ with purely imaginary
$\beta$.
This leads to very poor convergence in the numerical treatment of the
appendix and to check our results we repeated the calculation using a
simple
Taylor series basis (around $x=1/2$) for the wavefunctions and Richardson
extrapolation in the basis size.
This produced consistent results for small $\epsilon$, but a much larger
critical coupling $\epsilon_c \sim 0.75$.
It seems likely that $\epsilon_c$ is determined by the condition
${\rm Re~}\beta=0$ and solving for this condition in \bd\ numerically
produces
$\epsilon_c=0.676215 \pm 0.000005$.

As one decreases the quark mass, there will be an interesting and subtle
crossover between the behavior we just saw, and the case of a massless
meson,
which is unaffected by these perturbations.  It would be interesting to
know if
the resulting critical points are again roughly as expected from classical
considerations.  This requires more care in the numerics and we
reluctantly
leave the question for future work.

\newsec{Higher dimensions}

It is conceivable that these generalized actions could be of
direct value in higher dimensions.
First, we hasten to reassure the reader that in $D>2$,
expanding around the standard (free) UV fixed
point, the new terms $\tr E^n$ with $n>2$ are non-renormalizable.%
\footnote{*}{The only exception is the term
$\epsilon_{\mu\nu\lambda} \tr E^\mu[E^\nu,E^\lambda]$
in $D=3$.}
Rather than trying to define them in the continuum, we take as our
definition of generalized QCD$_D$ a lattice
gauge theory where for each
plaquette Boltzmann weight we use \weight.

The study of these generalized
theories in higher dimensions seems to go against two very widely accepted
principles.  The first is universality: the choice of bare action within
wide
limits is irrelevant to the study of the continuum theory, so why should
we not
be satisfied with the actions we know?  We would reply with a two
dimensional
analogy: there are many models whose long distance limit realizes the
physics of the three-state Potts model, for example,
but if one chooses to study Baxter's definition, one can find
exact solutions and make much quicker progress.
Although one is not optimistic about exact solvability in realistic
higher dimensional models, valuable technical simplifications might be
possible.

The second is that theories with higher derivative Lagrangians are
non-unitary.
We have already shown that this is not true in our generalized QCD$_2$ but
this
might be thought to be a special case, unitary because the gauge field has
no
local degrees of freedom.
However this is not the general explanation.
One element of the continuum explanation is that one can write the
corresponding continuum Lagrangians using higher powers of the field
strength
$F_{\mu\nu}$ but not higher derivatives of $F$, therefore preserving the
number of canonical degrees of freedom.

We now argue that generalized QCD$_D$ on a hypercubic lattice can satisfy
reflection positivity and then is the Euclidean continuation of a
unitary quantum field theory.
As is well known it suffices to have a positive, self-adjoint transfer
matrix.
Choosing a lattice axis as time, the transfer matrix could be taken as the
composition of an ``electric'' and ``magnetic'' operator,
\eqn\transone{T(\Delta t,a) = T_E(\Delta t,a) T_B(a)}
where $a$ is the spatial lattice spacing and $\Delta t$ the time step.
(We will suppress the $a$ dependence in the formulas below.)
We will take this to act on gauge-invariant wave functionals of the
spatial
link variables.

Clearly a more symmetric definition would be
\eqn\transtwo{T(\Delta t) = T_E(\Delta t)^{1/2} T_B T_E(\Delta t)^{1/2}}
and it would suffice to show that $T_E(\Delta t)^{1/2}$ and $T_B$ are
positive
and self-adjoint.
Since $T_B$ is simply a multiplication operator by the product of the
magnetic
Boltzmann weights, we need the individual Boltzmann weights to be
positive.
As for $T_E(\Delta t)^{1/2}$, in fact we can write
\eqn\hamone{T_E(\Delta t)^{1/2} = T_E(\Delta t/2) = \exp -{t H_E\over 2} }
for a self-adjoint Hamiltonian $H_E$.
This simply follows from the original definition of the generalized heat
kernel
action for the Hamiltonian $H$: we can show that
\eqn\transthree{{d\over dt} T_E(t) = -\sum_i H_i T_E(t)}
where $H_i$ is the Hamiltonian for the site $i$.
In detail:
we have
\eqn\tdef{(T_E(t)\psi)(U) = \int \prod_i dU_{ti} \prod_{ij} dU'_{ij}
\prod_{ij} Z_{at}(U_{ij} U_{tj}^{-1} U'_{ji} U_{ti}) \psi(U').}
Now the point is that
\eqn\point{{d\over dt} Z_t(U~V) = - H(E_U) Z_t(U~V)}
where the Hamiltonian is constructed from generators of left rotations of
$U$,
for all $V$.  We can therefore evaluate the r.h.s. of \transthree\ under
the integral.

We have shown that a generalized QCD$_D$ has reflection positivity (at
finite cutoff $a$),
if the solution of the generalized heat equation is positive
for all $U$ at time $a^2$.
The proviso is necessary for
positivity of the magnetic plaquette Boltzmann weights
and is not required in $D=2$.
That reflection positivity implies unitarity after continuation to
Minkowski space-time is clearest in  $A_0=0$ gauge, i.e. with the links
$U_{ti}=1$. This choice would simplify \tdef\ but have no effect on the
argument.

The condition of positivity of the solution of the generalized heat
equation
is quite non-trivial and in general is false.
Typically, positivity will require conditions like $t_k << t_2^{k-1}$
for $k>2$.  We have verified for the $U(1)$ case that there
are finite values of the couplings compatible with positivity and
have no reason to doubt it for $SU(N)$.
Even if this positivity is lost, it is not obvious that the Minkowski
theory
is nonsensical, because we always have a self-adjoint transfer matrix.
Correlation functions of operators whose time separations are even
multiples of
the lattice spacing will still satisfy reflection positivity.
\ref\parisi{G. Parisi, Statistical Field Theory,
Addison-Wesley, 1988; section 15.2.}
Whether this observation is relevant probably depends on the details of a
given
case.

A context in which a modified action has been proposed in the past is the
string interpretation of the
large $N$ limit of the strong coupling expansion
as derived in \refs{\kostov}.
The motivation can be illustrated in
$D=2$ and although we allude to \refs{\kostov} one could
make the same point starting from
the conceptually similar but simpler approach of Gross and Taylor \GT.
Without going into a detailed description of these results,
the basic idea is to expand the plaquette Boltzmann weight as
\eqn\plex{Z(U) = \sum_{\lbrace n_i\rbrace}\prod_i (N^i \Tr U^i)^{n_i} (1 +
O(1/N))
\exp -g^2 \left(\sum_i |n_i| + O(1/N)\right) }
(by rearranging the character expansion) and interpret a term
$\prod(\Tr U^i)^{n_i}$
as the contribution of a string configuration which locally has $n_i$
world-sheets with boundary wrapping $i$ times around the boundary of the
plaquette, in other words a $n$-fold cover of the plaquette with
$n=\sum_i |n_i|$.
The leading term in the exponential then is interpreted as a dependence
on the world-sheet area $A$ as $\exp -g^2 A$,
while the $O(1/N)$ terms (which can contribute to leading orders of the
free
energy)
are interpreted as due to insertions of
additional features at branch points or other singularities of the
covering.
This is where the modified actions we are considering become relevant --
changes to the action translate directly into changes to the $O(1/N)$
corrections of the action, and hence the additional world-sheet features.
The idea is that from the point of view of this expansion,
the simplest theory is not derived from the $\tr F^2$ action
but instead from an action which eliminates the $O(1/N)$ corrections
in the exponential.
In the language of \GT\ this corresponds to eliminating all world-sheet
features except for the $\Omega$ and $\Omega^{-1}$ points.
This theory has been discussed (to a limited extent) in \kostov\ and
\ref\douglas{M. R. Douglas, preprint RU-93-57, hep-th/9311130.}.

\nref\menotti{P. Menotti and E. Onofri, Nucl.Phys. B190 (1981) 288.}
The most useful descriptions of the action are either as a particular
case of \weight\ with $f(R)=\sum_i |n_i|$,
or as an expression similar to that of \menotti\ in which the eigenvalues
of
$U$
behave as free relativistic fermions and terms like $\exp -\theta^2/t$
are replaced by $1/(\theta^2+t^2)$.
In the form \first\ it apparently cannot be
expanded around $E=0$, the usual starting point for perturbation theory.
Nevertheless in the large $N$ limit a number of calculations can be done
with it and the results are non-singular.
In the calculations which can be reduced to free fermions, the explanation
is clearly that the fermi surface is at a `momentum' $E\sim N$ and we
never see the $E=0$ singularity.
However this is not a very general argument and one would like to see an
analogous phenomenon in calculations using other methods.
One of the original motivations for the present work was to examine this
question; however the results presented here do not suffice, as this
action requires products of the invariants $\tr E^n$ for its expression.
So far we see no reason why such models should be fundamentally harder to
solve, but we leave such questions for future work.

\newsec{Conclusions.}

In this work we have formulated a generalization of two-dimensional
gauge theory coupled to matter,
argued that the generalized models are renormalizable and unitary,
and using the techniques of the original work of 't Hooft,
solved a large subclass of such models in the large $N$ limit,
in the sense that we derived a integral equation which determines the
spectrum of mesons.
The simple classical picture of the linear potential produced by a quark
is qualitatively valid for the generalized interactions $\tr E^{2n}$, but
all generalized interactions produce additional short-range forces.

The models are qualitatively similar to the original model,
with a typical spectrum $m^2 \sim n$ for sufficiently massive states,
but can show new behaviors such as phase transitions or deconfinement.
This was illustrated in a preliminary study of the model
$\tr E F + \tr E^2 + \epsilon \tr E^3 + \bar \psi (\Dslash+m) \psi$.
The results also have bearing on some old questions, such as the validity
of
the principal part prescription as an infrared regulator
(in general, it is not).

We foresee two types of applications for these models.
The first is the possibility of choosing the action to produce a closer
analogy
to some feature of higher-dimensional gauge theory,
such as logarithmic violation of scaling.
The second, potentially quite important application, would be if a model
in
this large class turned out to have qualitatively similar physics to the
original model but was in some sense exactly solvable.
Perhaps most interesting would be a model which was not integrable in the
sense
of having a factorized S-matrix yet which allowed analytic calculation of
the
S-matrix to any order.
Presumably this could be done if closed expressions for the meson
wavefunctions
could be found.

Light-front quantization is of considerable interest as a non-perturbative
technique in higher dimensions, \who~
and these models provide new toy examples involving massless states,
symmetry breaking, and vacuum instability.

Given the interconnectedness of physics and more specifically
two-dimen\-sional
field theory, and the important role played by gauge fields,
it seems safe to predict that unforeseen applications will also be found.

It is also conceivable that these generalizations have some value in
higher
dimensions.  This is not the first time such a generalization has been
proposed, but we believe the present work significantly clarifies the
questions
which would need to be answered to justify the use of such an action.

\medskip
\leftline{\bf Acknowledgements}

We would like to thank T. Banks, P. Mende, N. Seiberg
and S. Shenker for valuable
discussions and W. Krauth for help on aspects of the numerical work.
M.R.D. acknowledges the support of DOE grant
DE-FG05-90ER40559, NSF grants PHY-9157016 and PHY89-04035, and the Sloan
Foundation. K.L. acknowledges the support of the W. M. Keck Foundation and
NSF grant PHY92-45317. M.S. acknowledges the support of
DOE grant DE-FG05-90ER40559 during the initial stage of this work.

\appendix{A}{Numerical methods}

Our numerical method is essentially the one developed in \num.
We expand the wave function in a basis of Chebyshev polynomials:
\eqn\basis{\eqalign{
\phi(x) &= \sum_{n\ge 1} c_n \phi_n(x)\hfill\cr
&= \sum_n c_n \sqrt{1-x^2} U_{n-1}(x)}}
where $U_n(\cos\theta) = \sin (n+1)\theta / \sin\theta$ and we have
redefined
$x$ to make the limits of integration $-1$ and $1$.
The weight factor is {\it not} the one which makes the basis orthonormal
but
rather was chosen so that we can apply a standard formula from the theory
of
integral equations
\ref\bateman{Higher Transcendental Functions, eds. A. Erd\'elyi et. al.,
McGraw-Hill, 1953; vol. 2, 10.11.48, p. 187.}:
\eqn\formula{
\int {dy\over y-x} (1-y^2)^{1/2} U_{n-1}(y) = -\pi T_n(x).}
This weight factor also gives the prescribed endpoint behavior
$\beta=1/2$ for the special case $\gamma=1$.
The basis is also quite suitable for $\beta>1/2$,
however it does a bad job at reproducing the endpoint behavior $\beta\sim
0$.
A reasonable cure for this problem (which we have not implemented here)
is to add another basis function such as $(1-x^2)^\beta$.

Combining with $T_n'(x) = n U_{n-1}$, and the orthogonality of the $U_n$'s
under $\int dx(1-x^2)^{1/2}$, we find that in this basis the 't Hooft
integral
has diagonal matrix elements $\pi^2 n/2$.
The inner product is
\eqn\inner{\eqalign{
(\phi_m,\phi_n) &= \half\int dx (1-x^2) U_{m-1}(x) U_{n-1}(x)\hfill\cr
&= \half\int_0^\pi d\theta sin\theta \sin m\theta\sin n\theta\hfill\cr
&= \cases{\half \left[ {1\over 1-(n-m)^2} - {1\over 1-(n+m)^2} \right]
	&n-m {\rm even};\cr
	0 &n-m {\rm odd}.}}}

The resulting generalized eigenvalue
problem is very well behaved numerically and
a
basis size of $100$ gives truncation errors less than $10^{-10}$ for the
low
lying masses.

Changing the quark mass requires the matrix element
\eqn\massint{\eqalign{
<m|{1\over 1-x^2}|n> &= \int dx U_{m-1}(x)U_{n-1}(x)\hfill\cr
&= \int  {d\theta\over\sin\theta}\sin m\theta\sin n\theta\hfill\cr
&=2\sum_{k=1}^{m-1} {1\over n+m-1-2k};\hfill\cr}}

We can adapt this to the higher kernels simply by computing their matrix
elements.
For an $E^3$ theory we would need
\eqn\hthree{\eqalign{
(\phi_m,H_3\phi_n) =&
i\int {dx dy\over (x-y)^2} \log\left({(1-x)(1+y)\over(1+x)(1-y)}\right)
 \tcbreak{\qquad} (1-x^2)^{1/2} U_{m-1}(x) (1-y^2)^{1/2} U_{n-1}(y).}}
The integrand now has only a single pole and the singular cutoff goes
smoothly
to principal part evaluation as $\lambda\rightarrow 0$.
This integral is particularly easy as we can write the log as the sum of
two
terms, and in (say) the $\log (1-x)/(1+x)$ term do the $y$ integral first,
using \formula\ and taking the derivative.
This will give
\eqn\hthreeint{\eqalign{
(\phi_m,H_3\phi_n) &=i\pi n \int dx (1-x^2)^{1/2} U_{m-1}(x)
\log\left({(1-x)\over(1+x)}\right) U_{n-1}(x) - (n \leftrightarrow
m)\hfill\cr
&= {2\pi} in \int_0^\pi d\theta \sin m\theta\sin n\theta
\log\cot{\theta\over 2} - (n \leftrightarrow m)}}
Using GR 4.384.7 we find
\eqn\grint{\int_0^\pi d\theta \cos m\theta\log\cot{\theta\over 2}
= \cases{{\pi\over |m|}&m {\rm odd}\cr 0&m {\rm even}}}
which applied to \hthreeint\ (keeping careful track of the absolute value
sign)
gives zero for $m-n$ even (as required by parity) and for $m-n$ odd gives
\eqn\hthreeval{
(\phi_m,H_3\phi_n) = i\pi^2\left({n-m\over|n-m|}-{n-m\over n+m}\right)
= 2\pi^2 i\ \sgn(n-m) {\min(n,m)\over n+m}.}

It is worth noting that the $i$ in the bound state equation can be removed
by a
unitary transformation (which symmetrizes \hthreeval).  Formally this
could be
done in \bs\ but it somewhat obscures the structure.

\listrefs
\end